\documentclass[twocolumn,prc,preprintnumbers,amssymb,
               floatfix]{revtex4}

\usepackage{dcolumn}
\usepackage{bm}
\usepackage{mathrsfs}
\usepackage{graphicx,epsfig,latexsym,amssymb}
\usepackage{multirow,amsmath,array,booktabs,color}
\usepackage[section]{placeins}
\usepackage{hyperref}
\usepackage{marvosym}
\usepackage{color}
\usepackage{epstopdf}
\usepackage{ulem}

\begin{document}
	
	\preprint{ }
	
	\title{Thermal properties of zero sound in asymmetric nuclear matter}

	\author{Jing Ye and Wei-Zhou Jiang\footnote{ wzjiang@seu.edu.cn}}
	\affiliation{  School of Physics, Southeast University, Nanjing 211189, China}

	\date{\today}
	
	\begin{abstract}
		The  zero-sound modes at finite temperature are investigated with the relativistic random phase approximation to signal the uncertainty of the equation of state (EOS) of asymmetric nuclear matter.  It is observed that  in typically selected stiff and soft  relativistic mean-field (RMF) models, zero-sound modes arise at low temperature, whereas increasing the temperature  gradually breaks the zero sound  in soft models, with a smaller density range compared to stiff models. At high density, the presence or absence of zero sound turns out to be correspondingly the character of the stiff or soft RMF  EOS. More strikingly, we find by analyzing the dispersion relation and sound velocity  that at finite temperature the zero-sound modes in RMF models with the stiff EOS undergo a  thermal bifurcation, resulting in the transform of zero sound into the first sound at some momentum $Q>T$. The thermally bifurcated sound branch in the stiff  models and the zero-sound branch in the soft models are both highly sensitive to the slope of the symmetry energy,  providing promising signals for the pending high-density symmetry energies.  In addition, it is found that there exists a nonlinear dispersion relation for  both the stiff and soft models that supports the zero sound in the relatively lower density region.
	\end{abstract}

	\maketitle

	\section{\label{sec:level1}INTRODUCTION}
	    The investigation of hot nuclear matter remains a vibrant field of research due to its intimate connection to fundamental astrophysical phenomena and nuclear physics issues, which include relativistic heavy-ion collision dynamics~\cite{li2008recent,sorensen2024dense,yong2022probing, wei2024} and stellar collapse processes with supernova explosions that may generate hot protoneutron stars~\cite{fryer2011gravitational, janka2012explosion,branch2017supernova}. A critical common element in these extreme environments created in the laboratories and celestial bodies is the presence of asymmetric nuclear matter, whose equation of state (EOS) fundamentally governs their behaviors and properties~\cite{oertel2017equations,glendenning2012compact, haensel2007neutron}.  The EOS of asymmetric nuclear matter  can  be thermodynamically characterized by the energy density's functional dependence on baryon number density, temperature, and isospin asymmetry.  It typically consists of the EOS of symmetric matter and the symmetry energy, while the latter represents  the energy difference per nucleon between pure neutron matter and symmetric nuclear matter. The EOS of   symmetric matter  is well constrained in the vicinity of  saturation density, but exhibits significant uncertainty and model dependence at high densities~\cite{lattimer2012nuclear,roca2018nuclear}. On the other hand, the symmetry energy and its slope have been well constrained to some average values at saturation~\cite{li2014topical,baldo2016nuclear}, albeit with the exception from the PREX II results~\cite{yue2022constraints}.   However, the density dependence of the symmetry energy at suprasaturation density remains poorly constrained~\cite{li2001isospin,oertel2017equations}. Various theoretical models can give rise to very different density profiles of the symmetry energy at high densities~\cite{li2008recent,xiao2009circumstantial,feng2010probing,neill2023constraining,reed2021implications}.

Not only do the uncertainties in the nuclear EOS affect the thermodynamic properties of asymmetric matter such as  the core-crust transition density of neutron stars~\cite{ducoin2011core,chomaz2004nuclear,horowitz2001neutron,li2023,zheng2012effects,wei2018crust},  liquid-gas phase transitions and phase boundaries~\cite{fukushima2010phase,zhang2013liquid,xu2015thermal,constantinou2015thermal,sharma2010nuclear,buyukcizmeci2005isospin,kumar2024modern}, but also can be conveyed into uncertain understandings of  the  properties of nuclei off $\beta$-stability ~\cite{jiang2010,danielewicz2014symmetry,wang2015properties,wang2015positioning,roca2018nuclear}, neutron stars~\cite{lattimer2016equation,steiner2005isospin,lattimer2004physics}, and the  dynamical evolution of massive stars including supernova explosions~\cite{page2012thermal,lattimer2001neutron}.  Since the constraints on the high-density EOS of asymmetric matter are  still rather limited, searching for new observables is of special significance. In our previous work, it was found that the zero-sound onset in nuclear matter is sensitive to the EOS stiffness at high densities~\cite{ye2023zero}.  As there were just very few discussions on the zero sounds at finite temperature in the literature, this work extends the previous study to hot nuclear matter,  in which we will check the in-medium behavior of zero sounds.  It was argued in the literature~\cite{kolomietz2001sound,shlomo2004hot} that the zero sounds would be converted into first sounds  under extreme condition. Exploring whether and how this conversion occurs remains an interesting and significant issue.

The zero-sound mode, first predicted by Landau, has been experimentally observed in Fermi liquids at zero temperature and later studied extensively in both neutron star matter and nuclear matter~\cite{chin1977relativistic,bedaque2003goldstone,aguilera2009superfluid,leinson2011zero,ye2023zero}. Following the  methodology established previously in Ref.~\cite{ye2023zero}, this work investigates the linear response of hot nuclear matter in the relativistic random phase approximation (RRPA) based on the relativistic mean-field (RMF) models~\cite{ma2001isoscalar,greco2003collective,lacroix2004nuclear,horowitz2001neutron,carriere2003low,li2023}. Employing the RRPA formalism, we determine the zero-sound modes through the zeros of the dielectric function and scrutinize  their thermal properties  in dense nuclear matter and their correlation with the nuclear EOS of asymmetric matter.   Interestingly, we find that the zero-sound modes in  hot nuclear matter may bifurcate at high densities with a strong correlation with the EOS stiffness and the slope of the symmetry energy.

The structure of the remaining paper is as follows. In
Sec.~\ref{sec:level2}, we briefly introduce the formalism of the relativistic mean-field models and the RRPA. Some derivations are given for the polarization functions and dielectric function in the RRPA equation. Sec.~\ref{sec:level3} presents numerical results and discussions  on the EOS of asymmetric matter, zero-sound modes at finite temperature, and the relationship between them. Much attention is paid to the extraordinary thermodynamic properties of the (zero-)sound modes at finite temperature.  At last, a summary is given in Sec.~\ref{sec:level4}.
	
	\section{\label{sec:level2}FORMALISM}
 The following Lagrangian with  the isoscalar-isovector $(\omega-\rho)$ coupling
term serves as the starting point~\cite{serot1992relativistic,bodmer1991relativistic,jiang2005charge}:
	\begin{eqnarray}\label{eq-L}
		\mathcal{L}=&& \bar{\psi}\left[i \gamma_{\mu} \partial^{\mu}-M+g_{\sigma} \sigma-g_{\omega} \gamma_{\mu} \omega^{\mu}-g_{\rho} \gamma_{\mu} \tau_{3} b_{0}^{\mu}\right] \psi \nonumber\\
		&&+\frac{1}{2}\left(\partial_{\mu} \sigma \partial^{\mu} \sigma-m_{\sigma}^{2} \sigma^{2}\right)-\frac{1}{3} g_{2} \sigma^{3}-\frac{1}{4} g_{3} \sigma^{4}\nonumber\\
		&&-\frac{1}{4} F_{\mu \nu} F^{\mu \nu}+\frac{1}{2} m_{\omega}^{2} \omega_{\mu} \omega^{\mu}+\frac{1}{4} c_{3}\left(\omega_{\mu} \omega^{\mu}\right)^{2}\nonumber\\
		&&-\frac{1}{4} B_{\mu \nu} B^{\mu \nu}+\frac{1}{2} m_{\rho}^{2} b_{0 \mu} b_{0}^{\mu}\nonumber\\
		&& +4 \Lambda_{V} g_{\rho}^{2} g_{\omega}^{2} \omega_{\mu} \omega^{\mu} b_{0 \nu} b_{0}^{\nu},
	\end{eqnarray}
where $\psi$, $\sigma$, $\omega$, and $b_0$ are the fields of the nucleon,
scalar, vector, and isovector-vector mesons, respectively. The  energy density $\epsilon$ and pressure $P$  with the RMF approximation can be derived based on the Lagrangian, and for their explicit expressions, refer to Refs.~\cite{ye2023zero,serot1992relativistic}.
As one part of the EOS of asymmetric matter,  the symmetry energy of the RMF models at zero temperature is given as~\cite{jiang2010}
\begin{equation}\label{eqsym}
    E_{sym}=\frac{1}{2}\left(\frac{g_\rho}{m_\rho^*}\right)^2 \rho_B
    +\frac{k_F^2}{6E_F^*}=\frac{1}{2\alpha}g_\rho b_0
    +\frac{k_F^2}{6E_F^*},
\end{equation}
where $m_\rho^*$ is the $\rho$-meson effective mass with
$m^*_\rho=\sqrt{m_\rho^2+8\Lambda_{\rm v}(g_\omega g_\rho\omega_0)^2}$, $\alpha$ is the
isospin asymmetry  with $\alpha=(\rho_n-\rho_p)/\rho_B$, and $E_F^*$ is the Fermi energy.  The slope  of the
symmetry energy at saturation density $\rho_0$ is defined as
\begin{equation}\label{eqslp}
L=3\rho_0\left.\frac{\partial E_{sym}}{\partial\rho_B}\right
|_{\rho_0}.
\end{equation}

Note that the thermal excitation in the Dirac vacuum leads to the modifications to the vector and scalar nucleon densities as
\begin{eqnarray}
  \rho_V&=&\rho_B=\sum_{i=p,n}\frac{2}{(2\pi)^3}\int d^3k (f_i(E_k^*)-\bar{f}_i(E_k^*)),\nonumber \\
  \rho_S &=&\sum_{i=p,n}\frac{2}{(2\pi)^3}\int d^3k \frac{M^*}{E^*_k} (f_i(E_k^*)+\bar{f}_i(E_k^*)),
\end{eqnarray}
where  $M^{*}=M-g_{\sigma}\sigma_{0}$ is the nucleon effective mass, $E^*$ is the effective energy $E^*_k=\sqrt{\boldsymbol{k}^2+M^{*2}}$, and $f_i$ and $\bar{f}_i$ are the nucleon and anti-nucleon distribution functions, respectively. Specifically, $\bar{f}_i(E_{k}^*,\nu_i)=f_i(E_k^*,-\nu_i)$, where $\nu_i$ is the effective chemical potential defined by
$\nu_i=\mu_i+\Sigma_i^0=\mu_i-(g_\omega \omega_0 +t_3 g_\rho b_0)$,
with $t_3=1$ and $-1$ for proton and neutron, respectively.

   In the RRPA approach, the Dyson’s  equation  for  the longitudinal polarization $\Pi_L$ can  be  written  as  a  matrix  equation
 \begin{equation}\label{eqrpa}
		\tilde{\Pi}_{L}={\Pi}_{L}+{\Pi}_{L}D_{L}\tilde{\Pi}_{L},
	\end{equation}
 in nuclear matter. The  lowest-order polarization is  a  $3\times3$ matrix
 \begin{equation}
		\Pi_{L}=\left(\begin{array}{ccc}
		
            \Pi_{s}^{n}+\Pi_{s}^{p} & \Pi_{m}^{p} & \Pi_{m}^{n} \\
			\Pi_{m}^{p} &-\frac{q^2}{q_\mu^2} \Pi_{L}^{p} & 0 \\
			\Pi_{m}^{n} & 0 & -\frac{q^2}{q_\mu^2} \Pi_{L}^{n}
		\end{array}\right),
	\end{equation}
where the individual polarization entries are given by
	\begin{subequations}
		\begin{align}
			i \Pi_{s}\left(q, q_{0}\right) &=\int \frac{d^{4} p}{(2 \pi)^{4}} \operatorname{Tr}[G(p) G(p+q)] ,\\
			i \Pi_{m}\left(q, q_{0}\right) &=\int \frac{d^{4} p}{(2 \pi)^{4}} \operatorname{Tr}\left[G(p) \gamma_{0} G(p+q)\right] ,\\
			i \Pi_{L}\left(q, q_{0}\right) &=-\frac{q_\mu ^2}{q^2}\int\frac{d^{4} p}{(2 \pi)^{4}} \operatorname{Tr}\left[G(p) \gamma_{0} G(p+q) \gamma_{0}\right],
		\end{align}
	\end{subequations}
	with Tr indicating the trace over Dirac indices. Here, the nucleon  Green function reads
\begin{align}
    G_{i}(k) = & \left(\gamma_{\mu}k^{\mu}+M^{*}\right)
    \left[ \frac{1}{k_{\mu}^{2}-M^{* 2}+i\varepsilon} \right. \nonumber\\
    & \left. + \frac{i \pi}{E_{k}^{*}} \delta\left(k_{0}-E_{k}^{*}\right) f_i(E_k^*) \right. \nonumber\\
    & \left. + \frac{i \pi}{E_{k}^{*}} \delta\left(k_{0}+E_{k}^{*}\right) \bar{f}_i(E_k^*) \right], \quad i=p,n.
\end{align}

While the real part of the polarization function is associated with the RRPA energy density of nuclear matter~\cite{ji1988nuclear} reflecting the excitation energy of resonance modes, the imaginary part of the polarization is linked to the reaction cross section~\cite{caillon1995relativistic,horowitz1991neutrino}.
As an example, we give the real and imaginary parts of $\Pi_L^i (i=n,p)$  at finite temperature explicitly as follows
 \begin{widetext}
\begin{equation}
\begin{aligned}
 \Re \Pi_L^i(q,q_0)=&\int_{0}^{\infty} \frac{\bar{k}}{\pi^2 }\left\{\frac{1}{2 \bar{k} q}\left(E_k^2-\frac{\left(q_\mu^2+2 E_k q_0\right)^2}{4 q^2}\right) \ln \left|\frac{q_\mu^2+2 E_k q_0+2 \bar{k} q}{q_\mu^2+2 E_k q_0-2 \bar{k} q}\right|\right. \\
 &+\frac{1}{2 \bar{k} q}\left(E_k^2-\frac{\left(q_\mu^2-2 E_k q_0\right)^2}{4 q^2}\right) \ln \left|\frac{q_\mu^2-2 E_k q_0+2 \bar{k} q}{q_\mu^2-2 E_k q_0-2 \bar{k} q}\right|
 \left.+\frac{q_\mu^2}{q^2}\right\} \cdot (f_i(E_k^*)+\bar{f}_i(E_k^*)) d E_k,
 \label{eq:RE}
 \end{aligned}
\end{equation}

\begin{equation}
\begin{aligned}
\Im \Pi_{L}^i(q,q_0)=&\frac{1}{8 \pi} \int_{E_d}^{\infty} \frac{q_\mu^2}{q^3 } \cdot\left(4 E_k^2+4 q_0 E_k+q_\mu^2\right)\left(f_i(E_k^*)+f_i(E_{k+q}^*)-2 f_i(E_k^*) \cdot f_i(E_{k+q}^*)\right) d E_k \\
&+\frac{1}{8 \pi} \int_{E_d^{'}}^{\infty} \frac{q_\mu^2}{q^3 } \cdot\left(4 E_k^2-4 q_0 E_k+q_\mu^2\right)\left(\bar{f}_i(E_k^*)+\bar{f}_i(E_{k+q}^*)-2 \bar{f}_i(E_k^*) \cdot \bar{f}_i(E_{k+q}^*)\right) d E_k,
  \label{eq:IM}
 \end{aligned}
\end{equation}
\end{widetext}
with $\bar{k}=|\boldsymbol{k}|$ . The lower limits in the integration for the imaginary part of the polarization function are obtained as
\begin{equation}
\begin{aligned}
&E_d=-\frac{q_0}{2}+\frac{q}{2}\sqrt{1-\frac{4M^{*2}}{q_\mu^2}},\\
&E_d^{'}=\frac{q_0}{2}+\frac{q}{2}\sqrt{1-\frac{4M^{*2}}{q_\mu^2}}.
 \end{aligned}
\end{equation}
 At zero temperature, the distribution function $f_i(E_k^*)$  is replaced by the step function $\theta(k_{F_i}-\bar{k})$. The expressions for $\Pi_s^i$ and $\Pi_m^i$ at zero temperature can be found in Ref.~\cite{lim1989collective}.

The interaction matrix $D_L$ in Eq.(\ref{eqrpa}), corresponding to meson propagators,  is given as follows~\cite{ye2023zero,li2024}
\begin{equation}
		D_{L}=\left(\begin{array}{ccc}

			\chi_{\sigma} & 0 & 0 \\
			0 & \tilde{\chi}_{V}+2 \tilde{\chi}_{\omega \rho} & \tilde{\chi}_{I} \\
			0 & \tilde{\chi}_{I} & \tilde{\chi}_{V}-2 \tilde{\chi}_{\omega \rho}
		\end{array}\right),
	\end{equation}
where the specific expressions of the matrix elements can be found in Ref.~\cite{ye2023zero}. 
 The collective excitation mode arises,
 when  the  dielectric function, which appears in the denominator of the RRPA polarization,   vanishes, i.e.
 \begin{equation}\label{Eq-epsilon}
		\varepsilon_L (q,q_0)=\operatorname{det}\left(1-D_{L} \Pi_{L}\right)=0.
	\end{equation}
This excitation mode turns usually out to be  zero sound at small momenta $q$, as long as it follows the dispersion relation that has the zero-energy  limit for vanishing momentum~\cite{ye2023zero}. Here, we give the explicit expression for the dielectric function:
 \begin{widetext}
\begin{align}\label{eqeL}
\varepsilon_L = & \left[1 - \left(\Pi_s^p + \Pi_s^n\right) \chi_\sigma\right]  \left[1 + \left(\Pi_L^n + \Pi_L^p\right)\left(\chi_\omega + \chi_\rho\right) \right. \notag  \left. + 4 \Pi_L^p \Pi_L^n \chi_\omega \chi_\rho + 2\chi_{\omega\rho}(\Pi_L^p - \Pi_L^n - 2\chi_{\omega\rho}\Pi_L^p\Pi_L^n) \right] \notag \\
& - \chi_\sigma \left[{\Pi_m^{p}}^2\left(\tilde{\chi}_\omega + \tilde{\chi}_\rho + 4 \tilde{\chi}_\omega \chi_\rho \Pi_L^n + 2\tilde{\chi}_{\omega\rho} - 4\tilde{\chi}_{\omega\rho}\chi_{\omega\rho}\Pi_L^n\right) \right. \notag  \left. + {\Pi_m^{n}}^2\left(\tilde{\chi}_\omega + \tilde{\chi}_\rho + 4 \tilde{\chi}_\omega \chi_\rho \Pi_L^p - 2\tilde{\chi}_{\omega\rho} - 4\tilde{\chi}_{\omega\rho}\chi_{\omega\rho}\Pi_L^p\right) \right. \notag \\
& \left. + 2 \Pi_m^p \Pi_m^n\left(\tilde{\chi}_\omega - \tilde{\chi}_\rho\right) \right].
\end{align}
 \end{widetext}
The zeroes of the dielectric function, corresponding to  the poles of the RPA polarization function, are dominated by subtle balance or cancellation between the repulsive vector part and attractive scalar part. This is actually  the grounds for searching the EOS-sensitive properties of the zero sound in this work. In the past, we studied the relationship between the EOS and zero sound at zero temperature~\cite{ye2023zero}. The zero-sound modes were found to prefer nuclear matter with the stiff EOS~\cite{chin1977relativistic,lim1989collective,ye2023zero}.  It was verified in the literature~\cite{matsui1981fermi,caillon2001landau} that zero sound obtained with the RRPA is consistent with the one with Landau's kinetic theory. Physically, both approaches address the response of nuclear matter to the perturbation. The density fluctuation in Laudau's kinetic theory is actually the variation of the source term for nuclear potentials that dominate the particle-hole excitations in the RRPA we adopt in this work.

	\section{\label{sec:level3}RESULTS AND DISCUSSIONS}

\subsection{RMF equations of state}
In order to check the EOS dependence of the zero sound,  we choose four  well-fit   models NL3w03~\cite{ye2023zero}, GM1~\cite{glendenning1991reconciliation}, TM1w02~\cite{ye2023zero}, and FSUGarnet~\cite{chen2015searching}, the former and latter two of which belong roughly to stiff and soft EOSs at high densities.  Notably, NL3w03 and TM1w02 represent modified versions of the NL3~\cite{lalazissis1997new} and TM1~\cite{sugahara1994relativistic} models, respectively. The additional isoscalar-isovector ($\omega-\rho$) coupling term~\cite{horowitz2001neutron} is introduced in these modified models  to soften the symmetry energy, which  accordingly results in the reduction in the radius of neutron stars~\cite{horowitz2001prc}.  Parameters and saturation properties are listed in Table~\ref{tab:Models}  for all these parameter sets.
 \begin{table*}
		\caption{\label{tab:Models}Parameters and saturation properties for various parameter sets. The meson masses $m_i~(i=\sigma,\omega,\rho$), the incompressibility $K_0$,  and the symmetry energy $E_{sym}$ are  in units of MeV. The saturation density $\rho_0$ is in units of fm$^{-3}$. }
		\begin{ruledtabular}
			\begin{tabular}{ccccccccccccccc}
				
				&$g_{\sigma}$&$g_{\omega}$&$g_{\rho}$ &$m_{\sigma}$&$m_{\omega}$&$m_{\rho}$&$g_{2}$&$g_{3}$
				&$c_{3}$&$\Lambda_{V}$&$\rho_{0}$&$K_0$&$M^{*}/M$&$E_{sym}$\\ \hline

				NL3w03&$10.217 $ &$12.868 $  &$5.664 $ &$508.270 $ &$782.501 $ &$763 $ &$10.431 $ &$-28.890 $
				&- & 0.03&$0.148 $ &$272.56 $ &$0.60 $ &$31.8$\\

				GM1&$8.700$ &$10.603$  &$4.060$ &$500.000$ &$782.500$ &$763$ &$9.235$ &$-6.131$
				&- & -&$0.153$ &$300.28$ &$0.70$ &$32.5$\\		
	
				TM1w02&$10.029  $ &$12.614  $  &$5.277  $ &$511.198  $ &$783.000 $ &$770  $ &$7.233  $ &$0.618 $
				&71.31  &0.02 &$0.145  $ &$281.20  $ &$0.63 $ &$30.7$\\
				
				FSUGarnet&10.505  &13.700  &6.945 &496.939  &782.500  &763  &9.576 &$-7.207$
				&137.981  & 0.04338 &0.153  &229.63  &0.58  &30.9\\

			\end{tabular}
		\end{ruledtabular}
	\end{table*}

\begin{figure}[htbp!]
		\includegraphics[width=0.48\textwidth]{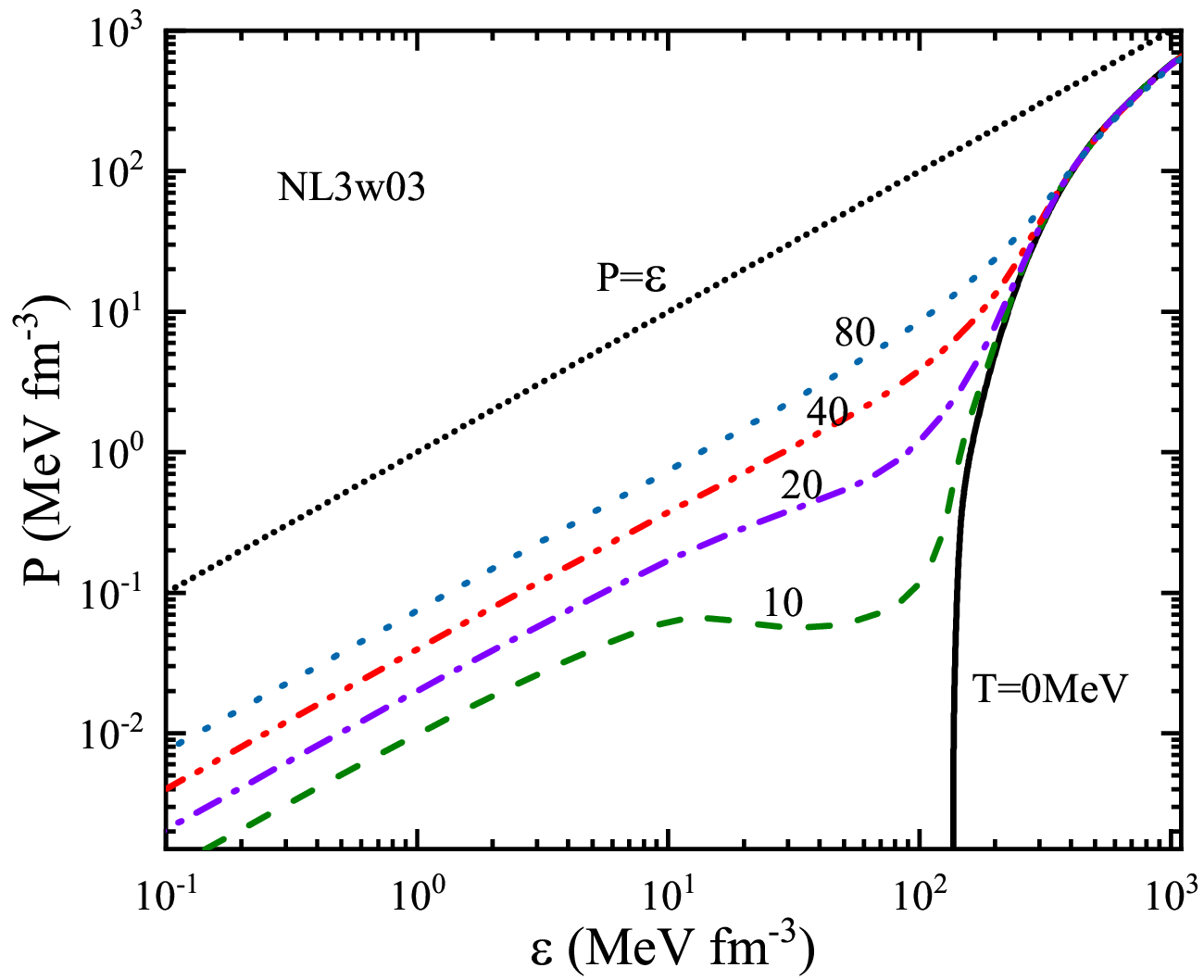}
		\centering
		\caption{\label{EOS-NL3w03} (Color online) Nuclear matter equation of state on isotherms with the NL3w03.  The black dotted line represents the causal limit, and the other curves represent the results at various temperatures as labelled.}
	\end{figure}

\begin{figure}[htbp!]
		\includegraphics[width=0.48\textwidth]{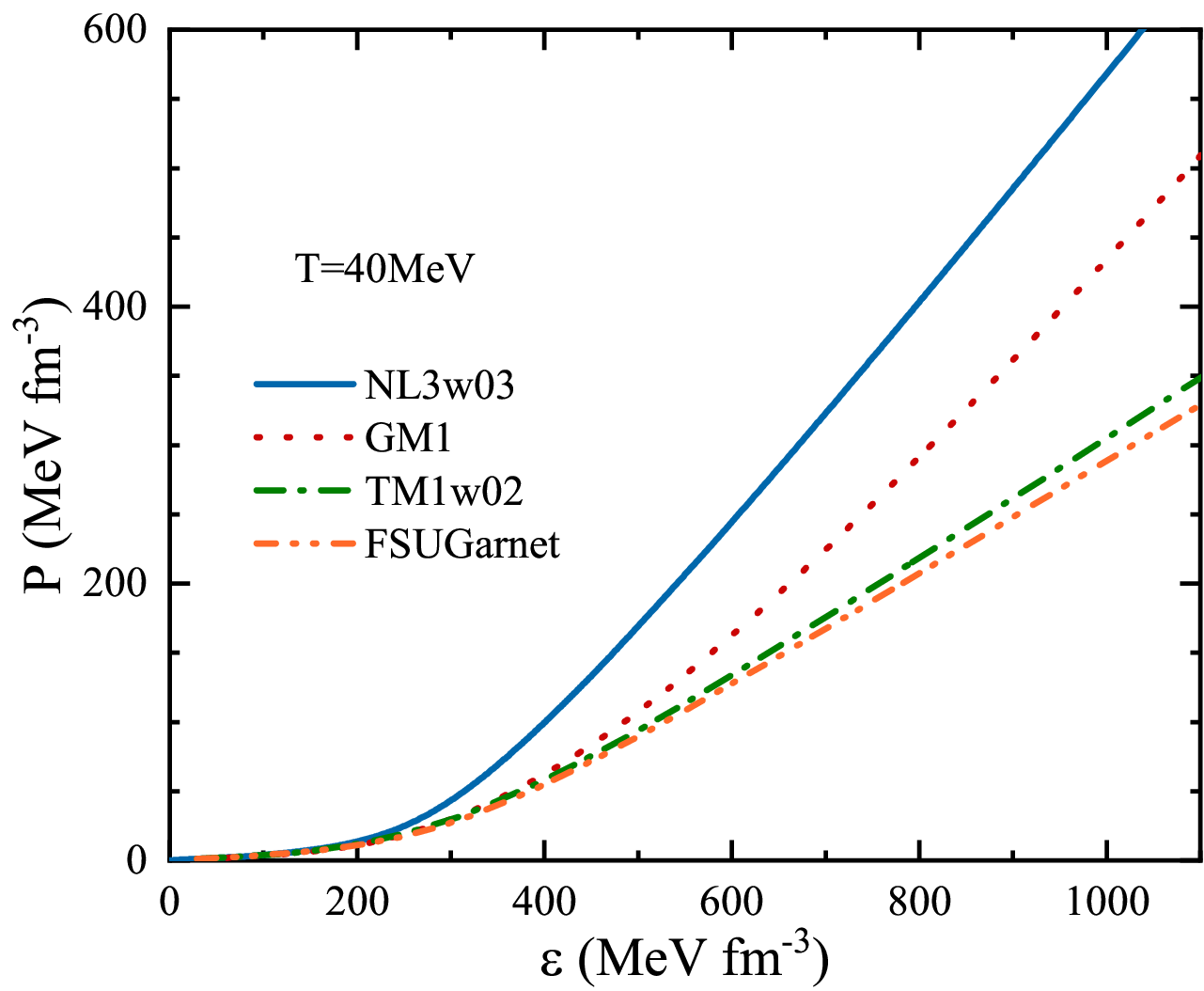}
		\centering
		\caption{\label{EOS-T} (Color online) The equation of state of symmetric nuclear matter with various RMF parameter sets, NL3w03,  GM1, TM1w02, and FSUGarnet {at temperature T=40 MeV.	} }
	\end{figure}

The EOSs for symmetric nuclear matter with parameter set NL3w03 at different temperatures are shown in Fig. ~\ref{EOS-NL3w03}.  Consistent with the result in the literature~\cite{serot1992relativistic}, the finite temperature  has  considerable effects on the EOS at low densities, instead of  a minor impact at high densities.  Taking the temperature T=40 MeV as an example, we illustrate the EOSs of the four   models in  Fig.~\ref{EOS-T}. It shows that the EOSs of the TM1w02 and FSUGarnet models are softer than those of the NL3w03 and GM1 models. Here, the softening is mainly due to the inclusion of the nonlinear self-interaction term of the $\omega$ meson ($\sim c_3$ $\omega_0^4$, see eq.~(\ref{eq-L})).

 	\begin{figure}[htbp!]
		\includegraphics[width=0.48\textwidth]{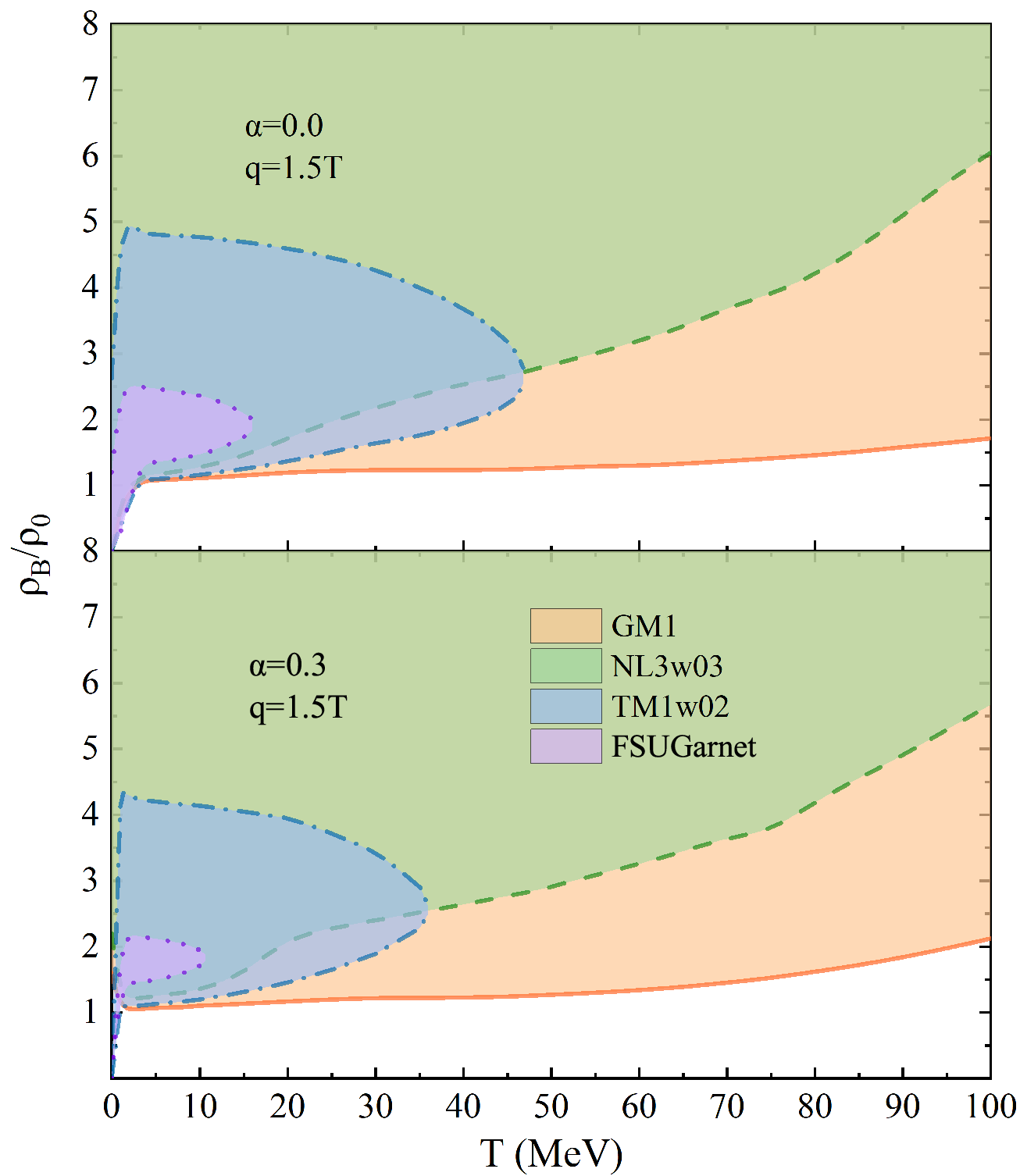}
		\centering
		\caption{\label{ZS-nB} (Color online) Contours of  zero sounds with various RMF models  for symmetric and asymmetric matter ($\alpha=0.3$) in the $T-\rho_B$ space. The momentum transfer ($q$) is set to 1.5 times the temperature ($T$). At zero temperature, $q$ is set to 1 MeV.	}
	\end{figure}

\subsection{Zero-sound occurrence and bifurcation at finite temperature}

In Fig.~\ref{ZS-nB}, we depict the contours of the zero sounds for symmetric and asymmetric nuclear matter in the temperature-density ($T-\rho_B$) space.  Following the equipartition theorem, the momentum transfer value ($q$) is set to be $1.5T$. It shows that the zero sound emerges over a broad $T-\rho_B$ region for stiff  EOSs, whereas the region narrows down noticeably for soft EOSs. Thus, the presence or absence of zero sound at quite high temperature and densities  turns out to be a clear marker for the stiffness of the EOS. In this high temperature and density regions, the occurrence of zero sound is primarily determined by the balance between vector repulsion and scalar attraction. On the other hand, the $T-\rho_B$ space for zero sounds in a rather large region reachable experimentally overlaps for the soft and stiff models. While the overlap is rather broad for the model TM1w02,  further EOS softening from TM1w02 to FSUGarnet can reduce the overlap region  appreciably. In contrast, such an overlap between the soft and stiff models does not occur at zero temperature~\cite{ye2023zero}. Indeed, the overlap is associated with the development of isovector and isoscalar zero-sound modes at finite temperature.  The isovector zero-sound modes at zero temperature, which appear at low densities in all models, extend to higher densities at finite temperature,while the isoscalar zero sound reaches out to  lower densities.  The mutual extensions of two modes in the $T-\rho_B$ space blur the boundaries for the isovector and isoscalar zero-sound modes at finite temperature, making it difficult to distinguish  the isoscalar and isovector zero-sound modes. Given a finite temperature, the expansion of the density range for zero sound is attributed to the intermediate-state contribution allowed by the lifted Pauli blocking.   At zero temperature, Pauli blocking strictly suppresses  the intermediate-state contributions, whereas with increasing temperature, the Pauli blocking is lifted by the thermal excitation that allows more intermediate-state contribution. This can be seen   in Eqs.~(\ref{eq:RE}) and~(\ref{eq:IM}) that  the step function at zero temperature is replaced by the finite-temperature distribution function, bringing the intermediate-state contributions and thermal   effects into effect.

As the high-temperature region is displayed in Fig.~\ref{ZS-nB}, it is necessary to discuss whether zero sound can persist in thermal collisions. Therefore, we examine the mean relaxation time between collisions $\tau$ at high temperatures, based on the kinetic theory of Landau-Fermi liquids~\cite{davison2012holographic,landau1981statistical,landau2013course,pines1966theory}.  The relaxation time  $\tau$ is approximated as~\cite{davison2012holographic} \begin{equation*}\tau \sim (1+e^{-1.5})\mu/({\pi^{2}T^{2}+1.5^2T^2}).
\end{equation*}
Using the model GM1, it is estimated at lower densities that $\tau = 4.6 \times 10^{-23} \text{s}$ at $ T=50 \text{MeV} $ ( $ \rho_B=\rho_0 $), and $\tau = 1.18 \times 10^{-23} \text{s}$ at $ T=100 \text{MeV} $ ($ \rho_B=1.5\rho_0 $ ), while at high densities it is  $\tau = 4.0 \times 10^{-23} \text{s}$ at $ T=50 \text{MeV} $ ( $ \rho_B=5\rho_0 $ ) and $\tau = 9.58 \times 10^{-24} \text{s}$ at $ T=100 \text{MeV} $ ($ \rho_B=5\rho_0 $).
These relaxation times are not so short, as compared to the time scale of strong interactions ($10^{-23}$ s).  It can be reasonably assumed that the thermal system is in the collisionless regime for zero sound, if the temperature is not very high, say, up to about $T=50$ MeV. We notice that the system at  temperature of 100 MeV would be  a typical collision regime with cautions to mention zero sound. Though zero-sound can survive as a quantum phenomenon at finite temperature,  it is actually  more fragile to thermal fluctuations at high temperature for the increasing $\Im\Pi_L$.

 	\begin{figure}[htbp!]
		\includegraphics[width=0.48\textwidth]{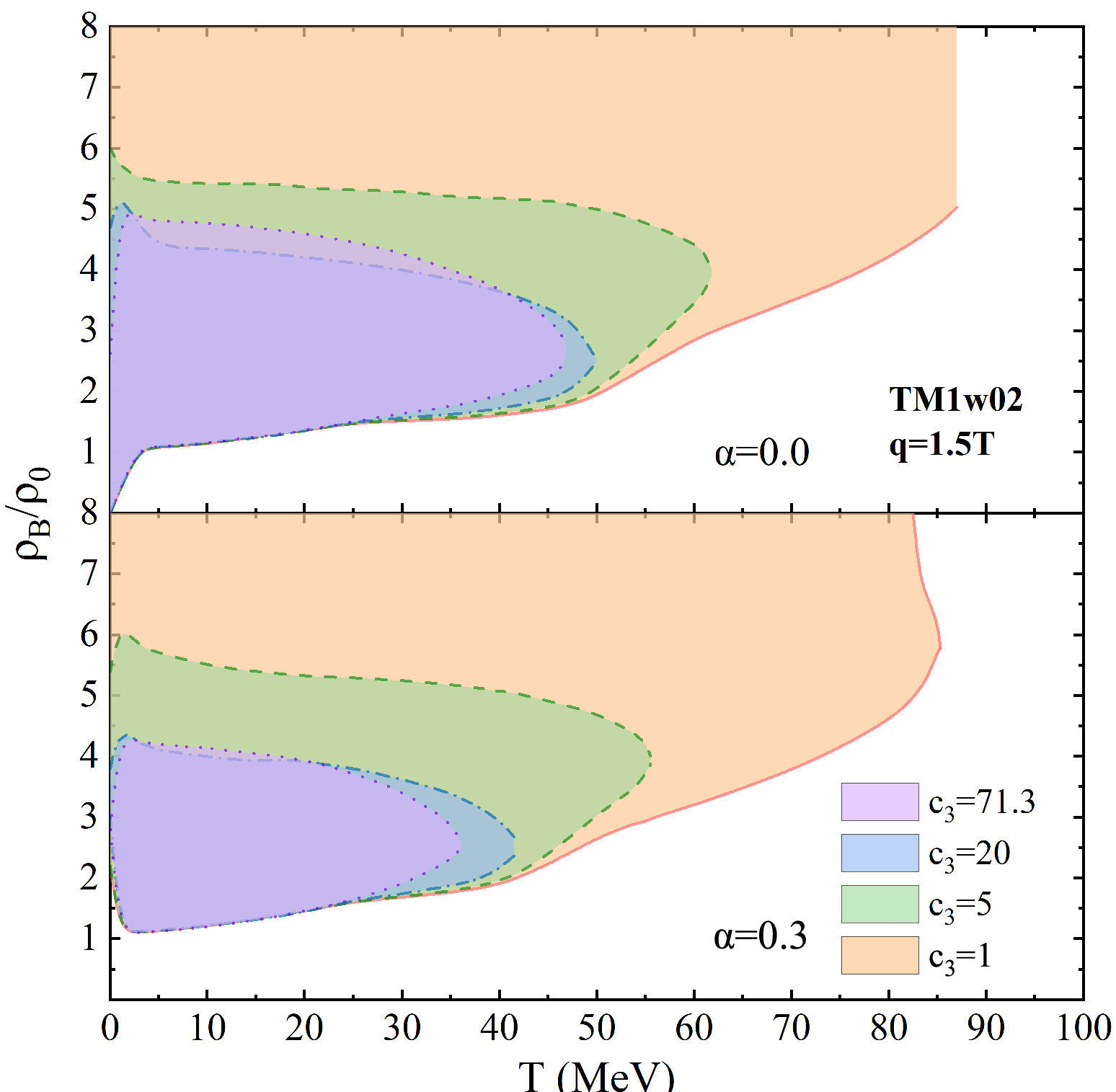}
		\centering
		\caption{\label{ZS-c3-TM} (Color online)  Contours of  zero sounds with various EOSs based on the TM1w02 for symmetric and asymmetric nuclear matter in the $T-\rho_B$ space.  The temperature and momentum  setups are the same as those in Fig.~\ref{ZS-nB}. 	}
	\end{figure}

To further examine the relationship between zero sound and the stiffness of the EOS at finite temperature, we adjust the coupling strength of the $\omega$ self-interacting term ($c_3$) to stiffen the soft TM1w02 model. The adjustment keeps the incompressibility unchanged, while other coupling constants in the model are moderately modified by less than 5\%~\cite{yang2017novel}. For readjusted  parameters, see Ref.~\cite{ye2023zero}.  Figure~\ref{ZS-c3-TM} shows that as $c_3$ decreases, corresponding to the stiffening of the EOS, the $T-\rho_B$ space, where zero sound appears in TM1w02, gradually expands to the regions of the high density and temperature ends.
Here, the expansion of the area for the appearance of zero sound in the $T-\rho_B$ space, due to the stiffening of the EOS, is similar to that in Fig.~\ref{ZS-nB}.  Consequently, whether the zero sounds occur at high temperature and density concerns the stiffness of the high-density EOS, which suffers the large uncertainty, and is involved in  understanding the cooling of neutron stars and the thermal evolution processes in heavy-ion collisions. On the other hand, a significant overlap of the zero-sound regions in the $T-\rho$ space appears at temperature up to about 25 MeV by varying the $c_3$ parameter in TM1.  The occurrence of this phenomenon is attributed mainly to the fact that  the region for the isovector zero-sound, which  is almost independent of the stiffness of the EOS at zero temperature, expands to the high-density region at finite temperature and melts mutually with  the isoscalar zero-sound mode that shifts downwards the lower density region.

\begin{figure}[htbp!]
\includegraphics[width=0.48\textwidth]{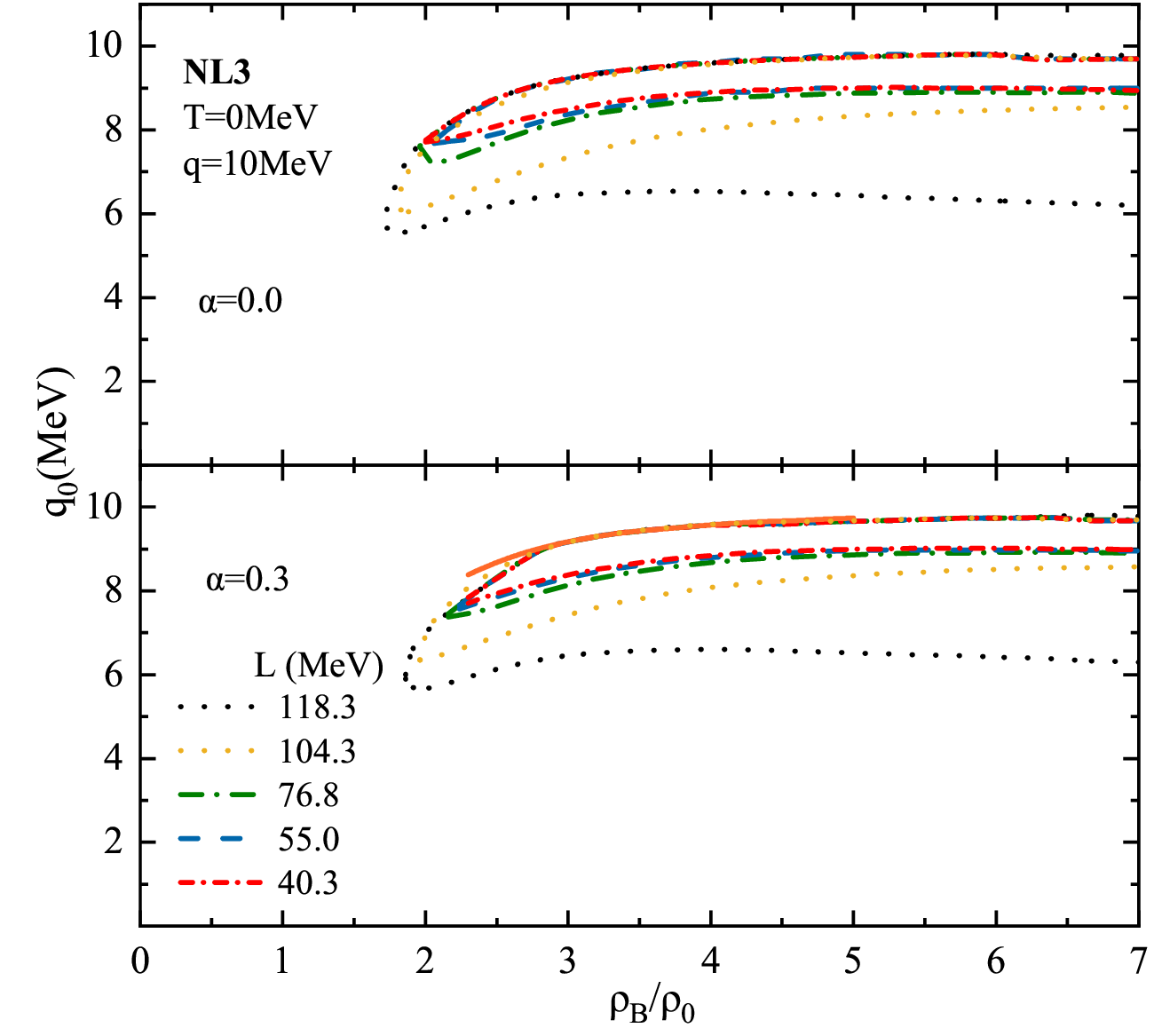}
		\centering
		\caption{\label{L-T0} (Color online)  Zero-sound modes in symmetric and asymmetric  ($\alpha=0.3$) nuclear matter  with various $L$ in NL3 at zero temperature and at $q=10$MeV.  In symmetric matter, the branch with the larger (smaller) $q_0$ corresponds to the undamped (damped) zero-sound mode. In asymmetric nuclear matter, three zero-sound branches are observed:  the undamped branch (solid curve)  and two damped branches. Note that the orange solid curve encompasses the overlapped regions of the undamped zero-sound modes with various slope parameters of the  symmetry energy.}
\end{figure}

\begin{table}
		\caption{\label{tab:L-NL3} Readjustment in  parameters  and symmetry energy in NL3 and TM1.   The symmetry energy $E_{sym}$ and the slope of the symmetry energy $L$  at zero temperature and saturation density $\rho_0$ are  in units of MeV. }
		\begin{ruledtabular}
			\begin{tabular}{ccccc}
				
				&$\Lambda_V$&$g_{\rho}$ &$E_{sym}(\rho_0)$&$L(\rho_0)$\\ \hline

				NL3&0.000  &4.474   &37.4  &118.3\\

				NL3w012&0.004 &4.854  &34.6 &83.0
    \\		
				NL3w015&0.015 &4.965  &34.1 &76.8
    \\		
	
				NL3w03& 0.030   & 5.664  &31.8 &55.0
    \\
				
				NL3w05&0.050  &7.323  &29.5 &40.3 \\
    \hline
				TM1&0.000  &4.632  &37.6 &113.1
  \\		
			   TM1w012&0.012  &4.988  &35.2 &82.4
    \\		

    			TM1w02&0.020  &5.277  &33.9 &68.2
    \\		
    			TM1w03&0.030  &5.720  &32.5 &55.2
    \\		
    			TM1w05&0.050  &7.095  &30.3 &39.1
    \\		
        \end{tabular}
		\end{ruledtabular}
	\end{table}

  	\begin{figure}[htbp!]
		\includegraphics[width=0.48\textwidth]{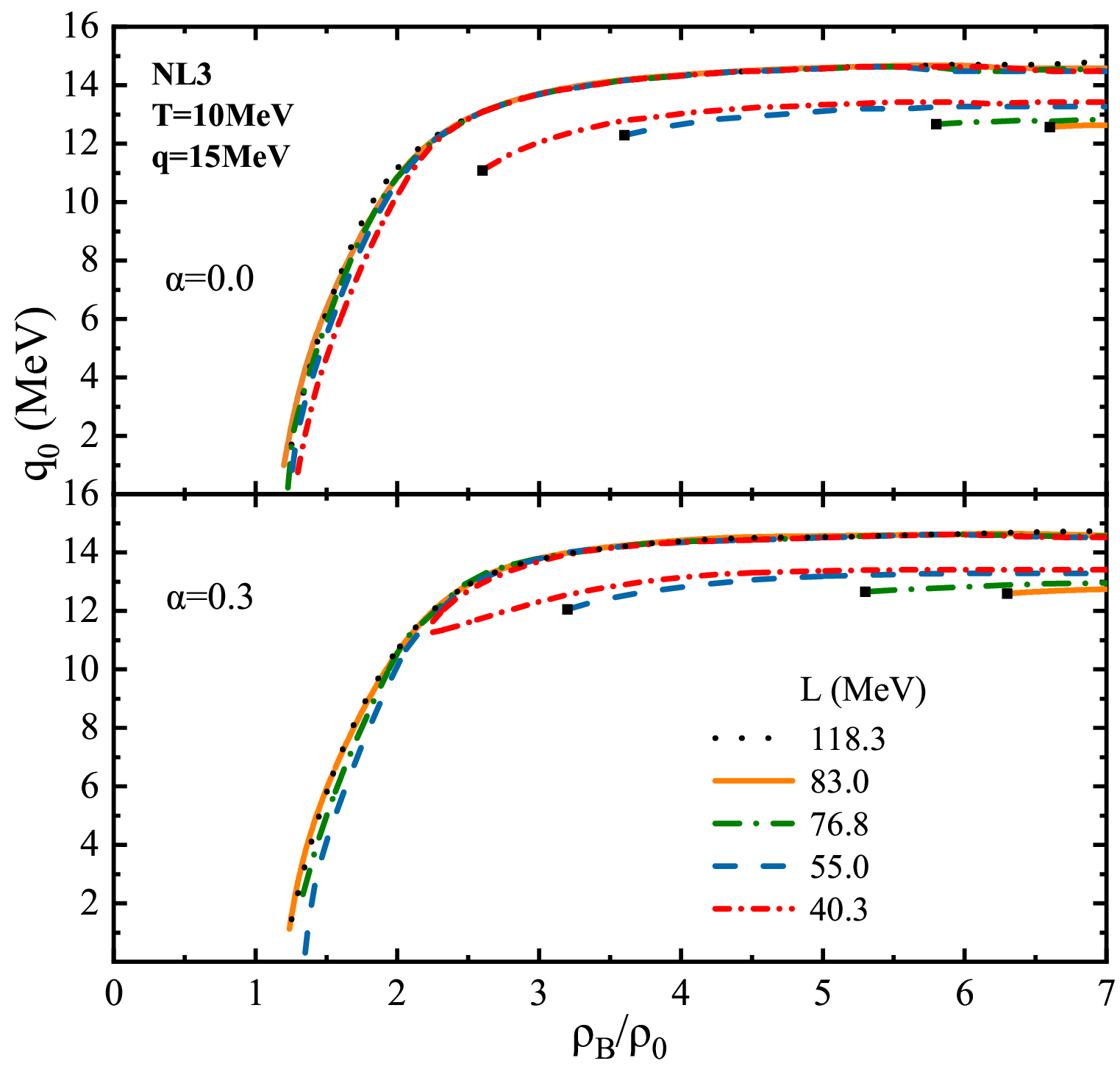}
		\centering
		\caption{\label{L-T10a0} (Color online) Density profile of (zero-)sound modes with various $L$ in NL3 at finite temperature. Here, $T=$ 10 MeV and $q=15$ MeV.  The black square represents the separate point away from the unphysical domain (not drawn) at lower densities.}
	\end{figure}

   	\begin{figure}[htbp!]
		\includegraphics[width=0.48\textwidth]{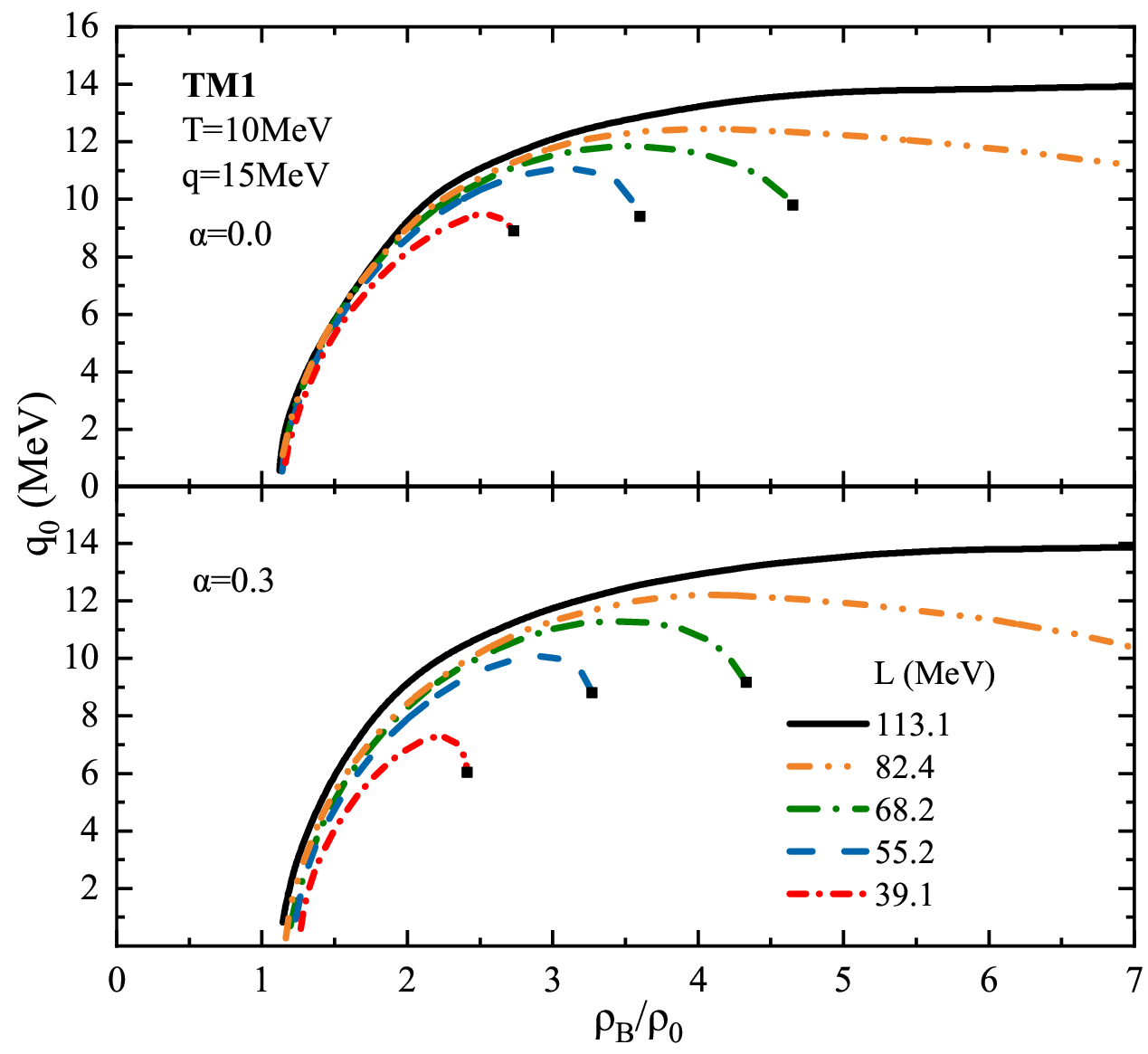}
		\centering
		\caption{\label{L-TM1} (Color online) Similar to  Fig.~\ref{L-T10a0} but using the TM1 model. }
	\end{figure}

Besides the EOS of symmetric matter, the symmetry energy, as one of the main contents of the asymmetric matter EOS, can also {affect the behavior of  zero-sound modes at both zero and finite temperature, since the interaction matrix $D_L$ possesses the isovector factor that can be modified by the symmetry energy.   To adjust the symmetry energy, we follow the method in Ref.~\cite{horowitz2001neutron} to modify  the isoscalar-isovector coupling  $\Lambda_V$ and the
$\rho$NN coupling constant $g_\rho$. As seen in Table~\ref{tab:L-NL3}, the symmetry energy in NL3 and TM1 is softened by increasing the value of $\Lambda_V$, together with a clear reduction in the slope parameter $L$.  At zero temperature, the energy of the damped zero-sound mode (the lower branches shown in Fig.~\ref{L-T0}) increases to approach  that of the undamped mode,  as the symmetry energy softens, both in symmetric and asymmetric nuclear matter. Meanwhile,  the undamped upper branch (with the larger $q_0$)  in symmetric  matter or the undamped (solid curve) and upper damped (broken curve) branches  in asymmetric  matter  remain essentially unchanged,  as shown in Fig.~\ref{L-T0}. This unchanged property is associated with the fact that  the polarization tensor $\Pi_L$ is sensitive to $q_0$. At large $q_0$, $\Pi_L$ can be quite small and the term $\chi_\rho\Pi_L$, which is much smaller than the isoscalar part due to the small isovector potential, can little affect the balance necessary for zero sound.

Figure~\ref{L-T10a0} shows the density profile of the zero sounds at finite temperature with the NL3 parametrizations.  Due to the influence of thermal fluctuations, zero-sound modes exhibit distinct characteristics  from those at zero temperature. In contrast to the case at zero temperature, both branches  exhibit damping characteristic ($\Im\Pi \neq 0$ in Eq.(\ref{eqeL}))  at finite temperature.
This is because thermal effects loosen the Pauli blocking and consequently smear out the zeros of the imaginary part of the polarization functions simply.  However, similar to the case at zero temperature, the low-energy branch of  zero sounds gets close to  the upper branch,  as the symmetry energy softens.  On the other hand, as the slope $L$ increases, the lower branch gradually recedes toward higher densities, as shown in Fig.~\ref{L-T10a0}. Once the $L$ exceeds 104.3 MeV in the NL3 parametrization, the lower branch is  missing across all density regions.  In contrast, we show in Fig.~\ref{L-TM1} the $L$ dependence of the zero sound in  the soft TM1 model.  It can be seen  that although the soft model TM1 features just one zero-sound branch,  distinct from the stiff model NL3, this unique branch remains sensitive to the symmetry energy: softening the symmetry energy by reducing $L$  causes the remarkable shrinkage of the zero-sound range  down to lower densities.   These results indicate that the symmetry energy can dramatically affect the collective modes, determined by the zeros of the dielectric function, but can have distinct behaviors in  the stiff and soft models.

 	\begin{figure}[htbp!]
		\includegraphics[width=0.48\textwidth]{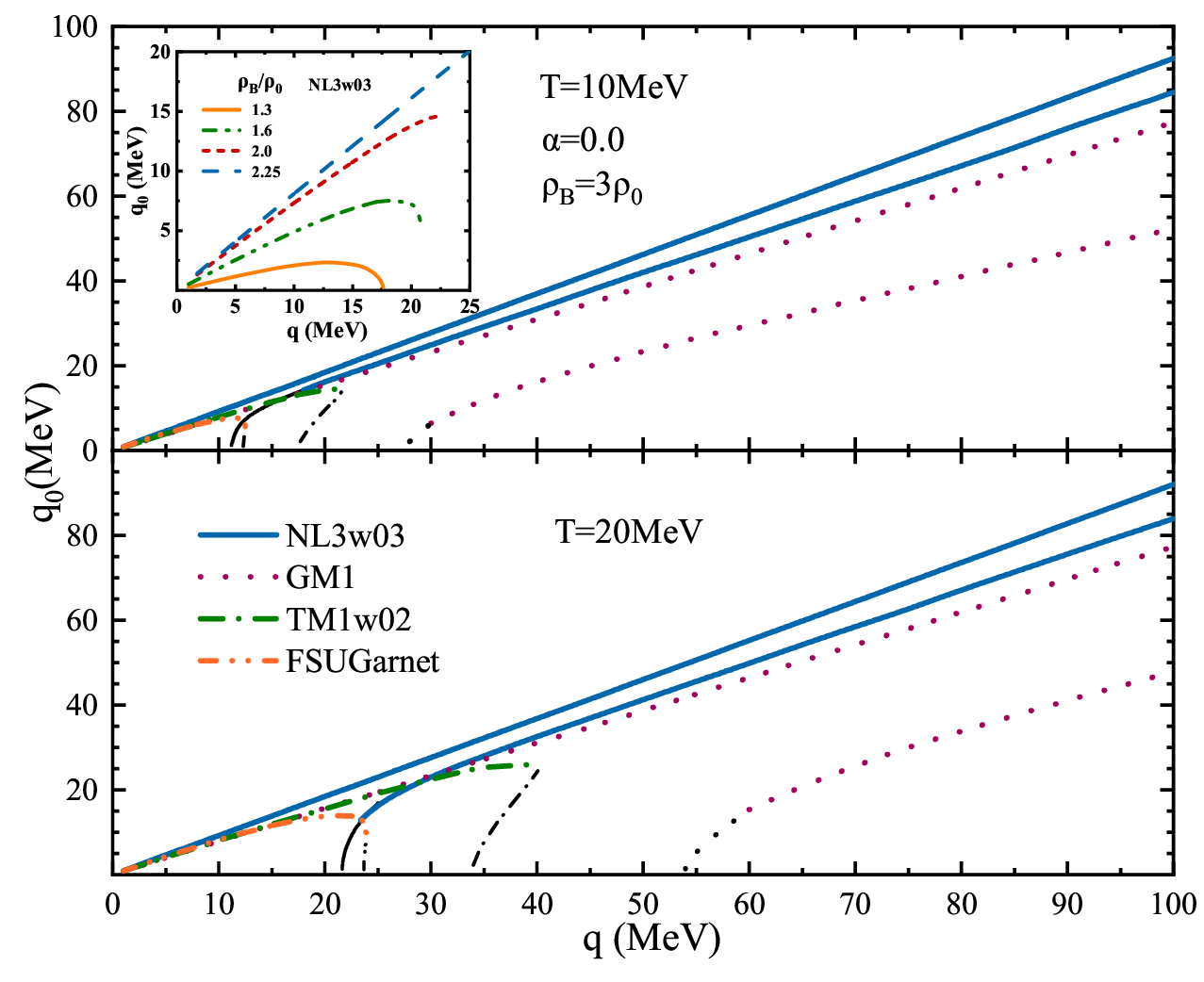}
		\centering
		\caption{\label{ZS-qq0} (Color online) Sound dispersion relation  in symmetric  matter at a density of $3\rho_0$ for various RMF models.  The upper and lower panels are at temperature $T=$ 10 MeV and 20 MeV, respectively. The  black curve reaches represent  the unphysical regions where the velocity exceeds the speed of light. The inset depicts the dispersion relation of the high-$q_0$ branch of zero sound  (see Fig.~\ref{L-T10a0})  at some lower  densities with NL3w03. }
	\end{figure}

The difference in zero sounds at zero and finite temperature, as compared in Fig.~\ref{L-T0} and Figs.~\ref{L-T10a0}, \ref{L-TM1}, ushers us to further examine the thermal characteristics and behaviors of zero sounds.  As a collective mode  determined by the zeros of the dielectric function (see Eq.~(\ref{Eq-epsilon})), zero sound should follow the specific dispersion relation that has the limit $q_0 \rightarrow 0$ for $q \rightarrow 0$.   At finite temperature,  the validity of this dispersion relation  maintains for zero-sound modes shown in Figs.~\ref{L-T10a0} (upper branches) and \ref{L-TM1}, corresponding to the dispersion relations in  Fig.~\ref{ZS-qq0} that pass through  the origin.
Shown in curves at the low-density end in Figs.~\ref{L-T10a0} and \ref{L-TM1}, the nontrivial behavior arises that given at $q=15$ MeV the energy of zero sound can approach to zero. This seems to deviate from the zero-sound dispersion relation. However, the dispersion relation is actually nonlinear in a certain density span and still has the limit $q_0\rightarrow 0$  by reducing the momentum $q\rightarrow 0$,  as shown in the inset in  Fig.~\ref{ZS-qq0}. Actually, the nonlinear dispersion relation at low density characterizes the instability of static matter~\cite{lim1989collective,fukushima2010phase,li2023strong}. As the density increases, $q_0$ does not approach zero at large q, ensuring a  well-defined low-energy limit for the zero-sound dispersion relation.
With increasing the density to $2.25\rho_0$, the nonlinear dispersion relation reduces to the linear one.

As shown in Fig.~\ref{ZS-qq0}, each parametrization gives two branches of the dispersion relation, and the lower branch   turns out to deviate from the zero-sound dispersion relation, that is, for the vanishing $q_0$, the momentum of the collective mode keeps away from the origin with a momentum stride $q>T$.
This departure is here termed the thermal  bifurcation of zero sounds which originates surely from the damped zero sound at zero temperature.
In addition,  the dispersion relations, shown in  Fig.~\ref{ZS-qq0}, exhibit distinct features specifically for soft and stiff models. For the soft models TM1w02 and FSUGarnet, the zero-sound and bifurcated branches appear only in the lower $q$ and $q_0$ regions at a given temperature and density, though increasing the temperature can expand the range of $q$ and $q_0$. In contrast, for the stiff models NL3w03 and GM1, the two branches exist over a much broader range of momentum and energy. It is worth noting that at finite temperature the thermally bifurcated  branch might exhibit spurious modes whose velocity exceeds the speed of light near $q_0=0$. It is verified that the bifurcated branch is totally spurious in the soft models TM1w02 and FSUGarnet, while it becomes  physical only for larger momenta in stiff models NL3w03 and GM1.  Actually,  the unphysical sound mode appears in   the space-like region ($q^2>q_0^2$) at finite temperature, which does not exclude the possibility of the  superluminal velocity of massless zero sound. One may think that  the temperature is not guaranteed  to be a Lorentz scalar~\cite{jiang2006relativistic}, and the thermal fluctuation  may perhaps affect the causality limit especially for  low-energy modes.  Alternatively, we can further analyze this phenomenon according to the Heisenberg uncertainty principle. At small $q_0$, the superluminal velocity stems from the small momentum difference $\Delta q$, which corresponds to a large space distribution of the zero sound. This would actually mean the melting of the zero sound. Since a quantitative interpretation is absent in our treatment, it is a pragmatic way  to classify the superluminal modes as unphysical ones.  Note that the unphysical domains are excluded in Figs.~\ref{L-T10a0} and \ref{L-TM1}.

A zero-sound branch still exists at finite temperature, which is somehow different from what was stated in Ref.~\cite{kolomietz2001sound} that at elevated temperature zero sound transitions to the first sound. However, a thermal bifurcation of the zero-sound modes in the stiff models results in a new branch  that is very sensitive to the symmetry energy. Since the bifurcated branch deviates from the zero-sound dispersion relation, it may transform to the first sound. In the following, we will explore how the bifurcated branch transforms into the first sound or  the relation between them by examining the velocity of  sound.

 \begin{figure}[htbp!]
		\includegraphics[width=0.48\textwidth]{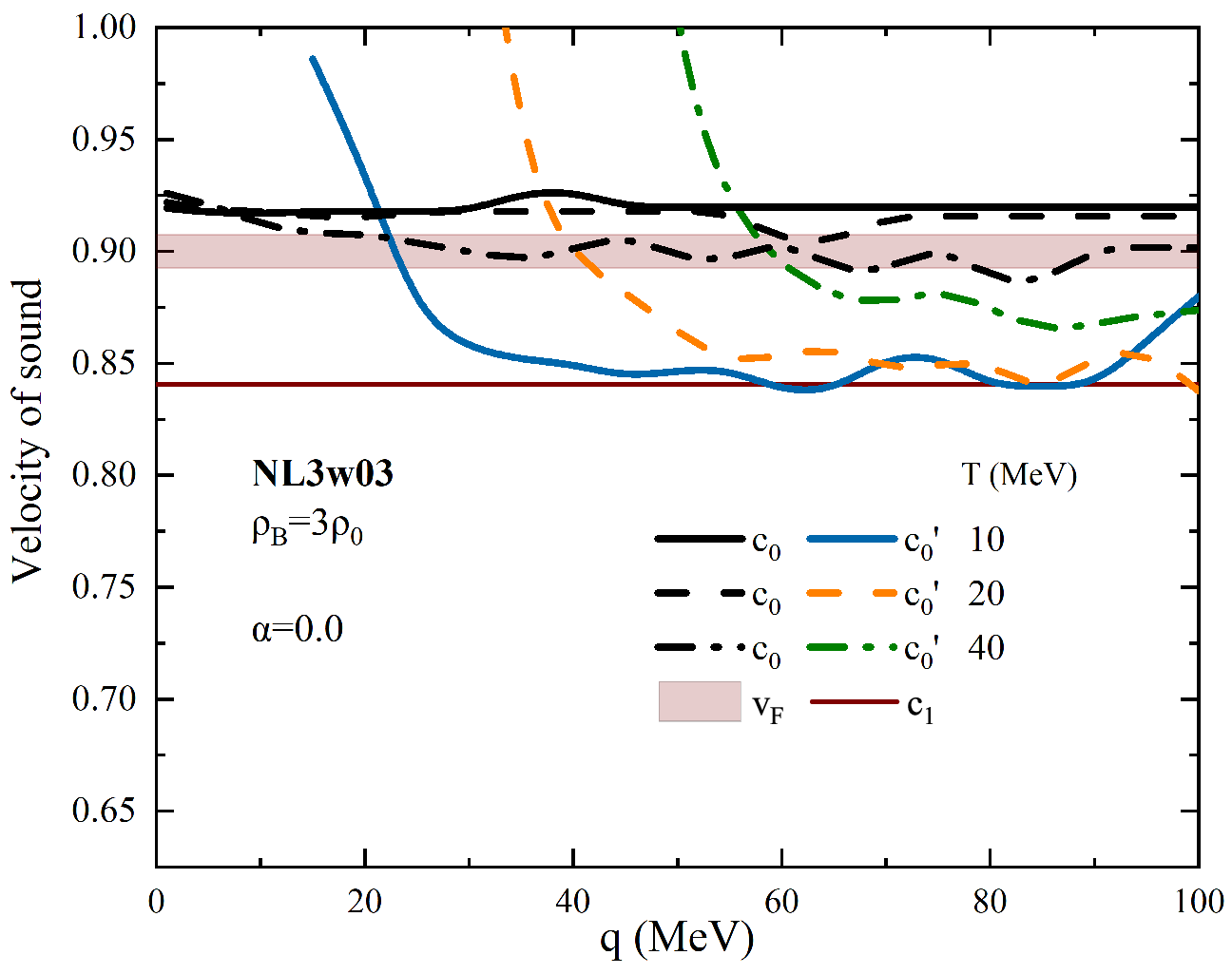}
		\centering
		\caption{\label{NL-T-qc0} (Color online) The  sound velocities as a function of the momentum $q$ at various temperatures with the NL3w03. The results are calculated at $\rho=3\rho_0$ in symmetric  matter. The black curves represent the velocity of  the zero-sound mode ($c_0$), while the colored curves represent the velocity of the thermally bifurcated sound ($c_0^\prime$).  The shaded area  stands for the $v_F$  range for temperature from $T = 10$ to 40MeV.	}
	\end{figure}

\subsection{Velocity of sounds}
We now investigate the velocity of sound at finite temperature. In comparison, we also calculate the speed of the first sound and the Fermi velocity so as to check their relationships.  The speed of the first sound  $c_1$  represents  the speed by which a longitudinal compression wave travels through medium. It is a fundamental quantity of  thermodynamic systems,  defined as $c_1=\sqrt{\partial P/ \partial  \epsilon}$ with  $P$ and $\epsilon$ being the matter pressure and energy density.  The speed of the first sound  reflects the EOS stiffness and thus affects crucially the phase transitions and evolutions of nuclear matter created by heavy-ion collisions~\cite{gardim2020revealing,sahu2020multiplicity,biswas2020viscosity, sorensen2021speed,he2023speed} and the neutron star properties such as  the mass-radius relation, cooling rate, maximum mass~\cite{ozel2016masses,zhang2019imprint,tews2018constraining,hebeler2013equation}. While the propagation of zero sound does not rely on the medium oscillation, it is interesting to look into the possible connection between the velocity of the first sound and bifurcated branch. Taking the model NL3w03 as an example, we show in Fig.~\ref{NL-T-qc0}  the various sound velocities  at different temperatures and at a specific density $\rho_B=3\rho_0$. Here, the speed of zero sound is defined by the relation $c_0=\partial q_0/\partial q$ along the dispersion relation.
As shown in Fig.~\ref{NL-T-qc0}, the velocity of the zero-sound branch is close to the Fermi velocity $v_F=k_F/E_F$, just showing a small variation depending on temperature.

In contrast,  the bifurcated sound exhibits  noticeable dependence on momentum ($q$). The physically bifurcated sound starts to appear only for some momenta $q=Q>T$.  At larger $q$ values, the velocity of the bifurcated sound shows a tendency  of being consistent with the speed of the first sound $c_1$.  This provides the evidence that the bifurcated sound branch transforms into the first sound. \ At finite temperature, the momentum excess ($Q>T$), which should be, strictly, a momentum domain having a lower boundary, provides a kinetic element  for the first sound in a plane wave form $\exp(iQr)$, whereas below the momentum $Q$ zero sound melts as discussed  above. Since thermal energy is assigned to quasiparticles at finite temperature,   the momentum excess has  a thermal origin and eventually drives the zero sound to transition to first sound. Instead of  a more quantitative interpretation, the transition to the first sound can be considered as  a nonlinear  emergence, kinetically driven by the thermal fluctuation.
As the temperature increases, say, to 40 MeV, the speed of the bifurcated sound gets a rise towards the Fermi velocity, as shown in Fig.~\ref{NL-T-qc0}. This behavior reflects likely the deficiency of the 1p1h excitation in the RRPA at high temperature.

	\section{\label{sec:level4}SUMMARY}
In this work, we have investigated the zero-sound modes at finite temperature in the RRPA based on the RMF models.  To search for the signals sensitive to the stiffness of the EOS,  the selected RMF  models are roughly classified into two categories with distinct stiffness of the EOS.    It shows that in hot nuclear matter both the stiff and soft RMF models produce the zero-sound modes at low temperature, whereas with increasing the temperature the soft EOS gradually ceases to produce zero sound. The $T-\rho_B$ space for zero sound in the soft EOS is also noticeably smaller than that in the stiff EOS. We further verify the dependence of the zero sound on the EOS by adjusting the EOS stiffness  in the RMF model.   When the soft RMF model is stiffened at high density by reducing the coupling strength of the $\omega$ self-interacting term ($c_3$), the density and temperature ranges for zero-sound modes  in both symmetric and asymmetric  matter expand gradually as expected. On the other hand, we find that there exists a universal nonlinear dispersion relation for  both the stiff and soft models that supports the zero sound in the relatively lower density region.

It is very striking to find by analyzing the dispersion relation and sound velocity  that at finite temperature the zero-sound modes in stiff RMF models  undergo a  thermal bifurcation that gives rise to a sound branch deviating from the zero-sound dispersion relation. At some momentum $Q>T$, the bifurcated sound branch shows a tendency to transform into the first sound. In contrast, the bifurcation and following transform into the first sound do not occur in  the soft EOS.   Noticeably,  the thermal bifurcated sound branch in the stiff EOS and the zero sound in the soft EOS are both highly sensitive to the slope of the symmetry energy at finite temperature.  As the slope of the symmetry energy reduces, the bifurcated sound branch in stiff models extends dramatically from high density to low density, whereas in soft models it is the density range for zero sound that recedes prominently  towards lower densities.
These features of zero sounds may have experimental implications in distinguishing the stiffness of high-density EOS.
Through intensive investigation and detailed comparisons with observational signals and experimental data, the correlation between zero sound and the EOS may assist us in better understanding the cooling of neutron stars and the thermal evolution processes in heavy-ion collisions.

	\section*{DATA AVAILABILITY}
Data will be made available on request.

	\section*{ACKNOWLEDGMENTS}
We thank J. Margueron and N. Li for helpful discussions. We are also grateful to the anonymous Reviewer for the valuable comments.
  This work was supported in part by the National Natural Science Foundation of China under Grants No. 12375112. The Big Data Computing Center of Southeast University is acknowledged for providing the facility support on
the partial numerical calculations of this work.

	\bibliography{ref}

\end{document}